\documentclass[epj,nopacs]{svjour}
\usepackage{graphics}
\usepackage{epsfig}
\usepackage{latexsym}
\usepackage{amssymb}

\newcommand{\spar}{{\stackrel{\rightarrow}{\Rightarrow}}}
\newcommand{\sant}{{\stackrel{\rightarrow}{\Leftarrow}}}

\begin{document}
\hugehead

\title{Double-spin asymmetries in the cross section of $\rho^0$
and $\phi$ production at intermediate energies.}
\titlerunning{Double-spin asymmetries in the cross section of $\rho^0$
and $\phi$ production.}
\authorrunning{The HERMES Collaboration}
\author{
The HERMES Collaboration \medskip \\
A.~Airapetian,$^{32}$
N.~Akopov,$^{32}$
Z.~Akopov,$^{32}$
M.~Amarian,$^{6,32}$
V.V.~Ammosov,$^{24}$
A.~Andrus,$^{15}$
E.C.~Aschenauer,$^{6}$
W.~Augustyniak,$^{31}$
R.~Avakian,$^{32}$
A.~Avetissian,$^{32}$
% ...
} 
\institute{ 
$^1$Department of Physics, University of Alberta, Edmonton, Alberta T6G 2J1, Canada\\
$^2$Physics Division, Argonne National Laboratory, Argonne, Illinois 60439-4843, USA\\
$^3$Istituto Nazionale di Fisica Nucleare, Sezione di Bari, 70124 Bari, Italy\\
$^4$Nuclear Physics Laboratory, University of Colorado, Boulder, Colorado 80309-0446, USA\\
$^5$DESY, Deutsches Elektronen-Synchrotron, 22603 Hamburg, Germany\\
$^6$DESY Zeuthen, 15738 Zeuthen, Germany\\
$^7$Joint Institute for Nuclear Research, 141980 Dubna, Russia\\
$^8$Physikalisches Institut, Universit\"at Erlangen-N\"urnberg, 91058 Erlangen, Germany\\
$^9$Istituto Nazionale di Fisica Nucleare, Sezione di Ferrara and Dipartimento di Fisica, Universit\`a di Ferrara, 44100 Ferrara, Italy\\
$^{10}$Istituto Nazionale di Fisica Nucleare, Laboratori Nazionali di Frascati, 00044 Frascati, Italy\\
$^{11}$Fakult\"at f\"ur Physik, Universit\"at Freiburg, 79104 Freiburg, Germany\\
$^{12}$Department of Subatomic and Radiation Physics, University of Gent, 9000 Gent, Belgium\\
$^{13}$Physikalisches Institut, Universit\"at Gie{\ss}en, 35392 Gie{\ss}en, Germany\\
$^{14}$Department of Physics and Astronomy, University of Glasgow, Glasgow G128 QQ, United Kingdom\\
$^{15}$Department of Physics, University of Illinois, Urbana, Illinois 61801, USA\\
$^{16}$Physics Department, University of Liverpool, Liverpool L69 7ZE, United Kingdom\\
$^{17}$Department of Physics, University of Wisconsin-Madison, Madison, Wisconsin 53706, USA\\
$^{18}$Laboratory for Nuclear Science, Massachusetts Institute of Technology, Cambridge, Massachusetts 02139, USA\\
$^{19}$Randall Laboratory of Physics, University of Michigan, Ann Arbor, Michigan 48109-1120, USA \\
$^{20}$Lebedev Physical Institute, 117924 Moscow, Russia\\
$^{21}$Sektion Physik, Universit\"at M\"unchen, 85748 Garching, Germany\\
$^{22}$Nationaal Instituut voor Kernfysica en Hoge-Energiefysica (NIKHEF), 1009 DB Amsterdam, The Netherlands\\
$^{23}$Petersburg Nuclear Physics Institute, St. Petersburg, Gatchina, 188350 Russia\\
$^{24}$Institute for High Energy Physics, Protvino, Moscow oblast, 142284 Russia\\
$^{25}$Institut f\"ur Theoretische Physik, Universit\"at Regensburg, 93040 Regensburg, Germany\\
$^{26}$Istituto Nazionale di Fisica Nucleare, Sezione Roma 1, Gruppo Sanit\`a and Physics Laboratory, Istituto Superiore di Sanit\`a, 00161 Roma, Italy\\
$^{27}$Department of Physics, Simon Fraser University, Burnaby, British Columbia V5A 1S6, Canada\\
$^{28}$TRIUMF, Vancouver, British Columbia V6T 2A3, Canada\\
$^{29}$Department of Physics, Tokyo Institute of Technology, Tokyo 152, Japan\\
$^{30}$Department of Physics and Astronomy, Vrije Universiteit, 1081 HV Amsterdam, The Netherlands\\
$^{31}$Andrzej Soltan Institute for Nuclear Studies, 00-689 Warsaw, Poland\\
$^{32}$Yerevan Physics Institute, 375036 Yerevan, Armenia\\
}  
\date{Received: \today / Revised version:}
\abstract{
Double-spin asymmetries in the cross section of electroproduction of $\rho^0$ and $\phi$ mesons on 
the proton and deuteron are measured at the HERMES experiment.
The photoabsorption asymmetry in exclusive $\rho^0$ electroproduction on the proton exhibits
a positive tendency. This is consistent with theoretical predictions that the exchange 
of an object with unnatural parity contributes to exclusive $\rho^0$ electroproduction by 
transverse photons. The photoabsorption asymmetry on the deuteron is found to be 
consistent with zero.
Double-spin asymmetries in $\rho^0$ and $\phi $ meson electroproduction by quasi-real photons 
were also found to be consistent with zero; the asymmetry in the case of the $\phi$ meson is compatible with a theoretical prediction which involves $s\bar{s}$ knockout from the nucleon.
}
\maketitle
\vspace*{-1.5cm}
\section{ Introduction}
\baselineskip=0.38cm
\vspace*{1.cm}

Traditionally, diffractive vector-meson production in lep\-ton-nucleon interactions is described as a 
fluctuation of the virtual photon into a quark-antiquark pair that subsequently forms 
a vector meson by scattering off the nucleon. 
For virtual photons with small negative four-momentum squared $Q^2<0.5$\,GeV$^2$, the formation of 
the $q\bar{q}$ state is usually described in the framework of Vector Meson Dominance 
(VMD)~\cite{vmd}, while at higher $Q^2$ it is assumed to follow the scheme of 
Generalised Vector Meson Dominance (GVMD)~\cite{vmd,gvmd}.
In terms of Regge phenomenology~\cite{regge}, the interaction of the virtual vector state with 
the nucleon can be explained  as an exchange of an intermediate object 
(Reggeon or Pomeron) in the t-channel of the reaction. 
For various vector mesons, different objects may be exchanged at different values of the invariant 
mass $W$ of the photon-nucleon system. 
In principle both Reggeon and Pomeron exchange can contribute to $\rho^0$ 
production, while in the case of $\phi$ production Reggeon-exchange amplitudes are expected 
to be strongly suppressed~\cite{freund}.

In the $Q^2$ range covered by the HERMES experiment both longitudinal and transverse photons
contribute to vector meson electroproduction~\cite{hermes_r,shilling,fraas_sdme}. For longitudinal photons 
information about the exchanged object can be extracted through cross-section 
measurements~\cite{hermes_cross}. For transverse photons this information is accessible 
through measurements of a double-spin asymmetry that arises in the cross section
and is sensitive to the parity\footnote{In principle, the parity of the 
object exchanged in vector-meson production can be also extracted from the full set of 
spin density matrix elements~\cite{shilling,fraas_sdme}.} of the exchanged object. 
No asymmetry can arise for longitudinal photons because their helicity is zero.

In general, the photoabsorption asymmetry $A^{}_1$ describing the spin dependence of the interaction between 
a transverse photon and a longitudinally polarised nucleon is defined as 
\begin{equation} 
A^{}_1=\frac{\sigma_{1/2}-\sigma_{3/2}}{\sigma_{1/2}+\sigma_{3/2}}\ .
\end{equation}
Here $\sigma_{1/2 \ (3/2)}$ stands for the transverse photoabsorption cross section where the 
subscript denotes the total helicity of the photon-nucleon system.
This asymmetry can be expressed in terms of the helicity amplitudes 
$T^{\lambda_{N'}\lambda_V}_{\lambda_N \lambda_\gamma}$, each of which receives contributions 
of both natural ($P=(-1)^J$) and unnatural ($P=-(-1)^J$) parities. Here $J$ denotes the total 
angular momentum of the exchanged particle and $\lambda_\gamma$, $\lambda_V$ and $\lambda_{N}$ 
($\lambda_{N'}$) indicate the helicity of the photon, vector meson and nucleon before 
(after) the interaction, respectively. In the approach of 
Ref.~\cite{fraas}, the asymmetry $A^{}_1$ arises from the interference between the parts of 
the transverse helicity amplitude $T^{11}_{11}$ with natural and unnatural parities. 
While a measurable asymmetry can arise even from a tiny contribution 
of the unnatural parity component, the latter may remain unmeasurable in the total cross section. 
A significant unnatural-parity contribution indicates the exchange of a di-quark or Reggeon. No 
asymmetry can be expected in the case of Pomeron exchange, since the Pomeron has natural parity.

The photoabsorption asymmetries presented in this paper are extracted with the aim of studying the 
mechanism of $\rho^0$ and $\phi$ production from transverse photons in the 
kinematic region covered by the HERMES experiment.
Double-spin asymmetries for the production of $\rho^0$ and $\phi$ mesons 
in lepton-nucleon scattering are presented, based on data 
obtained with a longitudinally polarised electron (positron) beam 
and longitudinally polarised hydrogen and deuterium targets.
Indication of a positive double-spin asymmetry in exclusive $\rho^0$ meson electroproduction 
on the proton was reported previously in Ref.~\cite{hermes_asym}. This asymmetry 
is here re-evaluated using an improved data set and a new parameterisation of $R$, the ratio 
of longitudinal to transverse photoabsorption cross sections. 

\section{ Experiment}

The HERMES experiment uses a target of polarised or unpolarised gas internal to 
the 27.5\,GeV electron (positron) beam of the HERA storage ring at DESY.
In 1996-1997 (1998-2000) the polarised target was operated with atomic hydrogen (deuterium). 
The lepton beam is transversely self-polarised by the emission of synchrotron radiation~\cite{st}. 
The longitudinal polarisation at the interaction point is obtained by spin rotators 
located upstream and downstream of the experiment. The beam polarisation is continuously measured 
by two Compton polarimeters~\cite{tpol,lpol}. The average beam polarisation for the proton (deuteron)
data set was 0.55 (0.55) with a fractional systematic 
uncertainty of 3.4 (2.0)\%.

The target~\cite{target} was fed by an atomic beam source, whose principle of operation is based 
on Stern-Gerlach separation in conjunction with hyperfine transition units. The average value of the target 
polarisation for the proton (deuteron) data set was 0.85 (0.85) with a fractional systematic 
uncertainty of 3.8 (3.5)\%.

The HERMES spectrometer is described in detail in Ref.~\cite{spectrometer}. Its 
angular acceptance in the laboratory frame spans the range $40<| \theta_y|<140$ mrad
and $|\theta_x|<170$ mrad, where $\theta_x$ and $\theta_y$ are the projections of the polar 
scattering angle into the horizontal and vertical planes.
The tracking system has a momentum resolution of about 1.5\%. The angular resolution is about 1 mrad.
Particle identification is accomplished by a lead-glass calorimeter~\cite{calo}, a preshower and 
a transition-ra\-dia\-tion detector. 
Until 1998 the particle identification system was complemented by a gas threshold 
{\v C}erenkov counter, which was then replaced by a dual-radiator ring-imaging 
{\v C}erenkov detector (RICH), described in detail in Ref.~\cite{richnim}. 
Combining the responses of these detectors in a likelihood method 
leads to an average electron (positron) identification efficiency of 98\% with a hadron contamination
less than 1\%.

\section{ Data analysis}
\subsection{ Kinematics}

At HERMES, a $\rho^0$ or $\phi$ meson is observed through its decay into 
two pions or kaons, respectively.
The kinematics of vector-meson production in lep\-ton-nuc\-le\-on scattering 
is described by $Q^2$, $W$, the energy $\nu$ of the virtual photon in the target rest frame
and the four-momentum transfer to the target $-t'=-(t-t_0)$, $t_0$ being the minimum longitudinal 
momentum transfer. 
The ``exclusivity'' variable $\Delta E=\frac{M^2_X-M^2_N}{2M_N}$ connects 
the mass of the target nucleon $M_N$ with the mass of the undetected hadronic system $M_X$. 
Also in case of the deuteron all kinematic variables were calculated using the mass of the proton.
The Bj{\o}rken scaling variable is defined as $x=Q^2/2\nu M_N$.

Two experimental topologies of vector-meson electroproduction at HERMES are considered in the following. 
The first one is denoted as 
{\it exclusive electroproduction}. Here the scattered lepton is detected in the spectrometer acceptance 
together with the meson decay products. 
The kinematics of the undetected recoiling nucleon can be reconstructed using those of the meson 
decay products and of the scattered lepton.
The exclusivity of the reaction is enforced by the requirement $\Delta E \approx 0$. 
At HERMES 
it results in the following average values: $\langle Q^2 \rangle$=1.8\,GeV$^2$, 
$\langle W \rangle$=4.9\,GeV, $\langle x \rangle$=0.07 and $\langle -t' \rangle$=0.15\,GeV$^2$.

The second topology is {\it electroproduction by quasi-real photons}. At low values 
of $Q^2$ the scattered lepton remains undetected in or close
to the beam pipe and the event kinematics cannot be 
fully determined from the data. 
The variable $\Delta E$ cannot be reliably calculated and hence 
exclusivity cannot be enforced for events of this topology. Due to the $Q^2$ dependence 
of the cross section, low-$Q^2$ events dominate those where the lepton is undetectable. 
The average values of $Q^2$ and $x$ for these events have been determined from Monte Carlo data, 
generated with the PYTHIA event generator\footnote{This generator was used for all Monte Carlo studies described in the paper except for the acceptance corrections.} version 6.1~\cite{pythia} 
tuned for the kinematics of HERMES. 
The photon structure was defined according to Ref.~\cite{friberg}. 
Candidate events for $\rho^0$ ($\phi$) meson electroproduction 
by quasi-real photons were selected requiring two accepted tracks belonging to oppositely 
charged pions (kaons) having a $\rho^0$ ($\phi$) as a parent particle. The average values of 
$Q^2$ and $x$ for these Monte Carlo events were calculated from the kinematics of the scattered positron 
that escaped the detector acceptance, resulting in $\langle Q^2 \rangle$=0.13 (0.12)\,GeV$^2$, $\langle W \rangle$=4.4 (4.2)\,GeV and $\langle x \rangle$=0.004 (0.006).   

\subsection{ Event selection}
The present analysis~\cite{lipka} of double-spin asymmetries in $\rho^0$ and $\phi$ meson production 
 is based upon data collected in 1996-2000, using longitudinally polarised 
hydrogen and deuterium targets. Candidates for exclusive $\rho^0$ meson 
($\phi$ meson) electroproduction were selected requiring exactly 3 tracks in the detector acceptance, 
corresponding to the scattered lepton plus two opposi\-te\-ly char\-ged pions (kaons). 
The vector-meson mass region was defined by the invariant mass 
constraint $0.6<M_{\pi\pi}<0.9$ GeV ($1.01<M_{KK}<1.03$\,GeV). Cuts were applied on the exclusivity 
parameter $\Delta E<0.6$\,GeV ($\Delta E<1.0$\,GeV) and the momentum transfer to the 
target $-t'<0.4$\,GeV$^2$ ($-t'<0.6$\,GeV$^2$). Note that in the case of the deuteron both coherent 
and incoherent parts of the vector-meson production cross section contribute at $-t'<$ 0.05\,GeV$^2$. 
The fraction of the events originating from coherent scattering is not negligible; the ratio of
coherent to incoherent cross sections was measured to be $0.160 \pm 0.015$.

 The photon energy was required to fall within 
$9<\nu<22$\,GeV. The lower limit is introduced by the kinematic relationship of $\nu$ and $\Delta E$, 
since at $\nu<9$\,GeV non-exclusive events occur at $\Delta E<3$ GeV, degrading the separation from 
the exclusive events. The upper cut ensures a high trigger efficiency. The $W$-acceptance of the HERMES
spectrometer for $\rho^0 \rightarrow \pi^+ \pi^-$ ($\phi\rightarrow K^+ K^-$) is sharply reduced both below 
4\,GeV and above 6\,GeV, and eliminates any contribution from the nucleon excitation region to $\rho^0$ ($\phi$) production.

Candidate events for $\rho$ ($\phi$) electroproduction by quasi-real photons were selected 
by requiring two tracks belonging to oppositely charged pions (kaons) in the detector 
acceptance. The same invariant mass constraints as in the case of exclusive electroproduction 
were applied.
 
For part of the data sample collected with the hydrogen target, hadron separation was accomplished 
with the {\v C}erenkov detector. The capability of this detector to identify pions is limited to 
the momentum range $p_h>3.5$\,GeV, which leads to losses in statistics. Therefore the information of 
the {\v C}erenkov detector was used only in the sample of electroproduction by quasi-real photons. In the data collected with the
deuterium target, hadron separation with the RICH detector was used. Restrictions on hadron
momenta were applied to provide efficient hadron identification~\cite{lipka}: $p_K>2$\,GeV, 
$p_\pi>0.5$\,GeV.

\subsection{ Extraction of double-spin asymmetries}

The formalism used here is described in more detail in Ref.~\cite{hermes_asym}.
The photoabsorption asymmetry $A^{}_1$ was extracted from the experimental
lepton-nucleon asymmetry $A^{}_{||}$ measured using a longitudinally po\-la\-ri\-sed lepton beam and target. 
These asymmetries are connected as follows:
\begin{equation}
\label{a1vm}
A_{1}^{} = \frac{A^{}_{||}}{D^{}}  - \eta \sqrt{R^{}}\ . \\
\end{equation}
Here $D^{}$ stands for the fraction of the beam polarisation carried by the photon and $\sqrt{R^{}}$ represents
the contribution~\cite{hermes_asym} from the asymmetry $A_2^{}$ arising from the interference between 
transverse and longitudinal photons, weighted by the small kinematic factor $\eta$,
where $R$ is the ratio of longitudinal to transverse cross sections. The definitions of these kinematic 
variables are given in Ref.~\cite{hermes_asym}.


In the calculation of asymmetries, background contributions have to be taken into account. 
Two main types of background can be distinguished: non-resonant back\-gro\-und from electroproduction of 
hadron pairs without formation of an intermediate vec\-tor-meson 
state, and non-exc\-lu\-si\-ve background from vector-meson production 
with the target nucleon not remaining intact.

The non-re\-so\-nant background can be subtracted by a fit to the invariant mass distribution, 
performed separately 
for each spin configuration of beam and target. 
The experimental asymmetry $A^{meas}_{||}$ is then calculated as follows:
\begin{equation}
\label{aparal_vm}
A^{meas}_{||}=\frac{1}{p_B \cdot p_T}\frac{N^\sant L^\spar-N^\spar L^\sant}{N^\sant L^\spar+N^\spar L^\sant} \ .
\end{equation}
Here $N^{\spar (\sant )}$ are the numbers of vector mesons produced with parallel (antiparallel) 
orientation of the nucleon helicity with respect to the lepton helicity. They are determined from the 
fitting procedure and are corrected here for the relative luminosity $L^{\sant (\spar )}$. The 
polarisations of beam and target are denoted by $p_B$ and $p_T$, respectively.

The asymmetry $A^{meas}_{||}$ still includes non-exclusive background events, which appear mostly at 
larger values of the exclusivity parameter $\Delta E$. This type of background is formed mainly by events whose final state 
contains a product of the fragmentation process in deep inelastic scattering (DIS), e.g. a hadron pair, 
a vector meson, or other particles decaying into them.
It is statistically impractical to fit the invariant mass distribution for each bin 
in $\Delta E$, in order to subtract non-exclusive background from each spin-dependent 
yield $N$. Therefore its contribution is taken into account as a dilution of the 
experimental asymmetry $A^{meas}_{||}$:
\begin{equation}
\label{aparal_exp}
A^{excl}_{||} = \ \frac{1}{1-r}\cdot(A^{meas}_{||} \ 
- r A^{ne}_{||}) \ .
\end{equation}
Note that for electroproduction by quasi-real photons the non-exclusive background can not be 
subtracted since neither the background asymmetry $A^{ne}_{||}$ nor the fraction $r$ can be determined from the data\footnote{From Monte Carlo data, the fraction of pure exclusive vector-meson electroproduction was obtained 
as 85\%, proton break-up occurs in 12\% of the cases and the remaining 3\% originate 
from other processes.}.

\subsection{Treatment of backgrounds}
\begin{figure}[ht]
\vspace*{-0.2cm}
\resizebox{0.5\textwidth}{0.35\textwidth}{\includegraphics{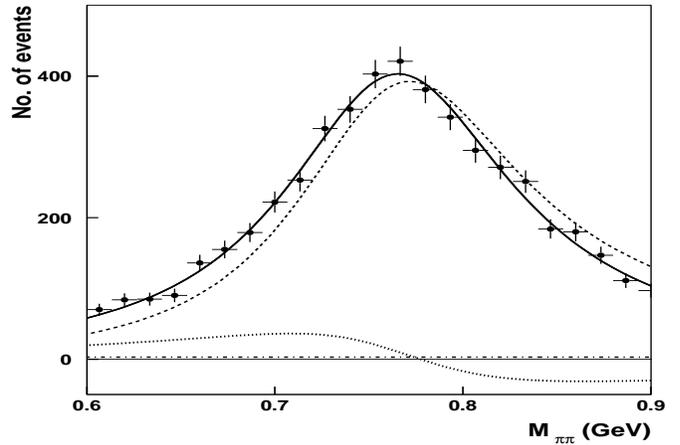}}
\caption{\baselineskip=0.38cm Subtraction of the non-resonant background. The invariant mass 
distribution, shown here for exclusive ($\Delta E<$ 0.6\,GeV) $\rho^0$ electroproduction on the deuteron, is fitted with a 
Breit-Wigner shape using the mass skewing model of Ref.~\cite{soeding} (solid line). The dashed 
line indicates the Breit-Wigner function, the dash-dotted line represents the non-resonant background, 
and the dotted line shows the interference term.}
\label{fig:1}   
\end{figure}

In the case of exclusive $\rho^0$ meson electroproduction, the pion pair invariant mass 
distribution was fitted with 
a relativistic $p$-wave Breit-Wigner function taking into account the skewing of the $\rho^0$ 
mass peak using the model of Ref.~\cite{soeding} (cf. Fig. \ref{fig:1}). The detector acceptance 
changes very little across the employed range in the invariant mass.

\begin{figure}[h!]
\hspace*{1.cm}
\resizebox{0.35\textwidth}{0.28\textwidth}{\includegraphics{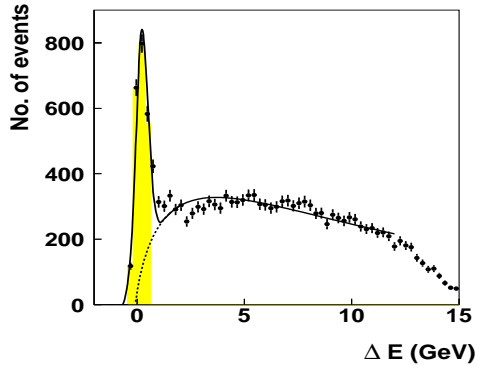}}
\caption{\baselineskip=0.38cm Subtraction of the non-exclusive background in $\rho^0$ 
electroproduction on the proton using a fit method. 
The $\Delta E$ distribution still includes the non-resonant background. The exclusive peak (shaded) 
is fitted by a Gaussian plus a background function (dashed line, cf. Eq. (\ref{rho_fun})). The solid 
line represents the sum of the Gaussian and the background.}
\label{fig:2}   
\end{figure}

The distribution of the exclusivity parameter $\Delta E$ is shown in Fig. \ref{fig:2}. 
The width of the exclusive peak  
is determined by the detector resolution, resulting in 0.28 (0.38) GeV for the detector 
configuration in 1996-1997 (1998-2000). Therefore some non-exclusive events appear also under 
the exclusive peak.

The asymmetry of the non-exclusive background, $A^{ne}_{||}$, was measured to be consistent with 
zero on both the proton and deuteron in the range 0.6\,GeV$<\Delta E<5$\,GeV. It is assumed that 
the asymmetry of non-exclusive events smeared into the exclusive region is the same.
The fraction $r$ of non-exclusive events in the exclusive region $\Delta E<$ 0.6\,GeV 
was estimated using a fit of an empirical function to the 
$\Delta E$ spectrum as shown in Fig. \ref{fig:2}. A Gaussian distribution was used for 
the exclusive peak. The background was described by the function 
\begin{equation}
f(\Delta E) = a_0 \cdot (\Delta E-a_1)\cdot e^{-a_2\sqrt{\Delta E-a_1}},
\label{rho_fun}
\end{equation}
where $a_0$, $a_1$ and $a_2$ are free parameters. This function is intended to account for 
all types of non-exclusive background like double-dissociative diffraction,
DIS and radiative tails. The part of the fitted background distribution 
falling within the exclusive region $\Delta E<$ 0.6\,GeV was taken as a measure of the 
non-exclusive background. The fraction of non-exclusive 
events was estimated to be $r=0.13\pm 0.01\pm 0.05$ (0.15$\pm0.01\pm0.07$) in the proton (deuteron) 
data set. The systematic uncertainty accounts for the range of background shapes that are 
compatible with the data, and is propagated into the systematic uncertainty of $A^{excl}_{||}$.

The spin-dependent yields in $\phi$ electroproduction were extracted in a way similar to those 
for $\rho^0$ production. 
\begin{figure}[h!]
\vspace*{-0.0cm}
\hspace*{-0.2cm}
\resizebox{0.5\textwidth}{!}{\includegraphics{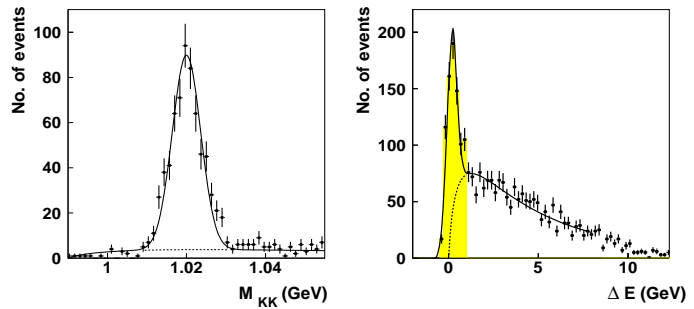}}
\caption{\baselineskip=0.38cm Subtraction of the non-resonant background (left panel) and of the non-exclusive background (right panel) in $\phi$ electroproduction on the deuteron. In the left panel the peak is fitted by a Gaussian (solid line) plus a background function 
(dashed line, cf. Eq. (\ref{fi_fun})). The solid and the dashed lines in the right panel 
have the same meaning as in Fig.~\ref{fig:2}. The shaded area indicates the exclusive region.}
\label{fig_fi}   
\end{figure}
Due to limited detector resolution the narrow invariant mass distribution of the kaon pair is widened 
with respect to its original shape. This effect was studied in Ref.~\cite{lipka}, 
using Monte Carlo events which were tracked through the detector 
taking into account the efficiencies and the resolutions of the detector subsystems. The 
smeared $\phi$ resonance peak in the invariant 
mass distribution of kaon pairs was described by a Gaussian (cf. Fig.~\ref{fig_fi}, left panel).
 For the background the empirical function 
\begin{equation}
f(M_{KK})=b_0\cdot (M_{KK}-M^{min}_{KK})\cdot e^{b_1\cdot\sqrt{M_{KK}-M^{min}_{KK}}}
\label{fi_fun}
\end{equation}
was used, where $b_0$ and $b_1$ are free parameters and $M^{min}_{KK}$ denotes the threshold of the kaon-pair 
invariant mass distribution, corresponding to the opening of phase space.
The fraction $r$ of non-exclusive events was estimated as described above for the case of $\rho^0$
production (cf. Fig.~\ref{fig_fi}, right panel) and was found to be $r=$0.28$\pm 0.03 \pm 0.10$ for 
both targets.

\subsection{ Acceptance corrections}
The data on exclusive electroproduction were corrected for acceptance effects. The corrections 
were obtained using a multi-dimensional look-up table, calculated from Monte-Carlo data in 
bins of $Q^2$, $x$, $-t'$ and $M_{\pi\pi}$, to account for possible correlations 
between these variables. The Monte Carlo simulations for the acceptance studies were 
performed using a generator~\cite{tytgat} based on the VMD model. 
The resulting acceptance correction factors were used as weights for every 
event depending on its kinematics. 
The effect of the acceptance correction manifested itself in a shift of at most 8\% of the 
original mean value of the bin center in $Q^2$ and $x$ at which the asymmetry was 
evaluated. No shift in $t'$ was observed.

\section{ Results and Interpretation}

The photoabsorption asymmetries $A_1$ for $\rho^0$  and $\phi$ electroproduction by quasi-real 
photons were determined directly from $A^{meas}_{||}$ using Eq. (\ref{a1vm}). 
In the case of exclusive electroproduction the data were additionally corrected for non-exclusive 
background using Eq. (\ref{aparal_exp}), i.e. the photoabsorption asymmetries $A_1$ were 
calculated from $A^{excl}_{||}$.
The ratio $R$ was extracted from the elements of the spin-density matrix for vector mesons produced 
at HERMES~\cite{tytgat,rakness}. 
The resulting values of $A_1$ averaged over the kinematics for both the proton and deuteron 
are listed in Table \ref{tab_asy}, together with the numbers of $\rho^0$ and $\phi$ mesons 
used in the analysis. 
For exclusive electroproduction of $\rho^0$ mesons, the asymmetries $A_1$
were calculated in several bins of $x$, $Q^2$ and $t'$. 
In Table~\ref{asy_kin_tab} the asymmetry values are shown
in dependence on each of the three kinematic variables while averaging over the other two. 
All asymmetries were found to be consistent with zero within experimental uncertainties, 
possibly apart from the asymmetry in exclusive electroproduction of $\rho^0$ mesons on the proton. 
In accordance with an earlier HERMES result~\cite{hermes_asym}, the latter was found to have a 
positive value, 1.7 $\sigma$ away from zero.

\begin{table}[h]
\begin{center}
\begin{tabular}{|l|l|l|}
\hline
\multicolumn{1}{|c|}{}& proton & deuteron\\
\hline
\multicolumn{3}{|c|}{Exclusive electroproduction}\\
\hline
$A^\rho_1$       & 0.23 $\pm$ 0.14$\pm$0.02 & -0.040 $\pm$ 0.076$\pm$0.013\\ 
$A^\phi_1$       & 0.20$\pm$0.45$\pm$0.03  & 0.17$\pm$0.27$\pm$0.02\\
$N^\rho$         &1774                   & 6505\\
$N^\phi$         & 219                   & 618 \\
\hline
\multicolumn{3}{|c|}{Electroproduction by quasi-real photons}\\
\hline
$A^\rho_1$  & 0.0057 $\pm$ 0.0093$\pm$0.0004  & -0.0039 $\pm$ 0.0029$\pm$0.0003\\ 
$A^\phi_1$  & 0.052 $\pm$ 0.084$\pm$0.003  & 0.018 $\pm$ 0.028$\pm$0.001\\
$N^\rho$    & $423\times10^{3}$ & $4013\times10^{3}$\\
$N^\phi$ & $7.6\times10^{3}$ &  $57\times10^{3}$ \\
\hline
\end{tabular}
\vspace*{0.3cm}
\caption{\baselineskip=0.38cm Photoabsorption asymmetries in vector-meson electroproduction and the numbers of $\rho^0$ and $\phi$ mesons used in this analysis. The statistical and systematic uncertainties of the asymmetries are given.}
\label{tab_asy}
\end{center}
\vspace*{-0.5cm}
\end{table}

The statistical uncertainties of the extracted numbers of vector mesons per helicity state and of the
fraction and asymmetry of the non-resonant background were propagated into the statistical uncertainty of 
the asymmetry $A_{||}$.  
The systematic uncertainties from the measurements of $A^{excl}_{||}$, of beam and target polarisation 
and the uncertainty from the parameterisation of $R$ (cf. Eq. (\ref{a1vm})) were combined to form 
the experimental systematic uncertainties to the asymmetry measurements. The systematic uncertainties 
are found to be considerably smaller than the statistical ones. 

\begin{table}[h]
\begin{center}
\begin{tabular}{|c|c|c|c|c|}
\hline 
                      &                     &                              &        \multicolumn{2}{c|}{}\\
\multicolumn{1}{|c|}{} & $\langle D \rangle$ & $\langle\eta\sqrt{R}\rangle$ & \multicolumn{2}{c|}{$A^\rho_1$} \\
\multicolumn{1}{|c|}{} &                     &                        &$^1$H           &  $^2$H \\
\hline
\hline
\multicolumn{1}{|c|}{$\langle x \rangle$}    & \multicolumn{4}{c|}{}  \\
\hline
     0.032      &  0.48           &  0.029             & 0.20 $\pm$ 0.20 & -0.13 $\pm$ 0.10 \\

0.057      &  0.34           &  0.055             & 0.28 $\pm$ 0.26 & 0.07 $\pm$ 0.14 \\

0.105      &  0.28           &  0.098             & 0.30 $\pm$ 0.34 & 0.09 $\pm$ 0.18 \\
\hline
\multicolumn{1}{|c|}{$\langle Q^2 \rangle$,\,GeV$^2$}  &  \multicolumn{4}{c|}{}  \\
\hline
 0.84  & 0.42         & 0.037              & 0.17 $\pm$ 0.19 & -0.03 $\pm$ 0.10 \\
 1.44  & 0.34         & 0.062              & 0.23 $\pm$ 0.26 & -0.16 $\pm$ 0.14 \\
 3.01  & 0.31         & 0.099              & 0.51 $\pm$ 0.36 & 0.19 $\pm$ 0.20 \\
\hline
\multicolumn{1}{|c|}{$\langle -t' \rangle$,\,GeV$^2$}    & \multicolumn{4}{c|}{}   \\
\hline
 0.018 & 0.36 & 0.059                     & 0.36 $\pm$ 0.30 & -0.09 $\pm$ 0.15 \\
 0.065 & 0.37 & 0.059                     & 0.16 $\pm$ 0.30 & -0.05 $\pm$ 0.15 \\
 0.138 & 0.36 & 0.062                      & 0.11 $\pm$ 0.31 & -0.11 $\pm$ 0.16 \\
 0.302 & 0.36 & 0.063                      & 0.46 $\pm$ 0.30 & 0.17 $\pm$ 0.16 \\
\hline 
\end{tabular}
\end{center}
\caption{\small Measured values of the photoabsorption asymmetry $A_1^\rho$, shown for various 
values of each of three kinematic variables while averaging over the other two. Total uncertainties are given.}
\label{asy_kin_tab}
\vspace*{-0.3cm}
\end{table}  

\begin{figure}[h]
\vspace*{-0.5cm}
\hspace*{0.3cm}
\resizebox{0.48\textwidth}{0.40\textwidth}{\includegraphics{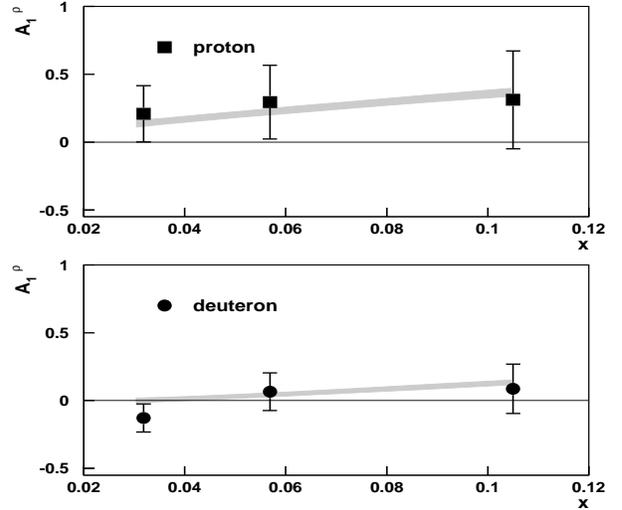}}
\caption{\baselineskip=0.38cm The $x$-dependence of the asymmetry $A^\rho_1$ in exclusive 
$\rho^0$ meson electroproduction on the proton (top) and deuteron (bottom). The data are compared 
to the expectations of Refs.~\cite{fraas,hermes_asym} expressed by Eq. (\ref{asy_expected}), 
as indicated by the shaded bands. The error bars represent the total uncertainties obtained by 
adding statistical and systematic uncertainties in quadrature.}
\label{fig:3}   
\end{figure}

The $x$--dependence of the photoabsorption asymmetry $A_1^\rho$ in exclusive $\rho^0$ electroproduction 
is shown in Fig. \ref{fig:3} for both the proton and deuteron. 
This measurement is compared to the expectation of Refs.~\cite{fraas,hermes_asym} that
 is based on a 
relation between the double-spin asymmetries $A_1^\rho$ in exclusive $\rho^0$ electroproduction and $A_1^N$ in inclusive DIS:
\begin{equation}
\label{asy_expected}
A^\rho_1  = \frac{2 A^{N}_1}{1+(A^{N}_1 )^2} \ .
\end{equation}
Using the HERMES measurements~\cite{a1_inc} of inclusive DIS asymmetries $A^N_1$, values for the expected 
asymmetries $A^\rho_1$ were calculated from Eq. (\ref{asy_expected}).
The measured asymmetries were found to be consistent with this expectation for both the proton and deuteron.
\begin{figure}[h!]
\vspace*{-1.cm}
\hspace*{0.3cm}
\resizebox{0.48\textwidth}{0.38\textwidth}{\includegraphics{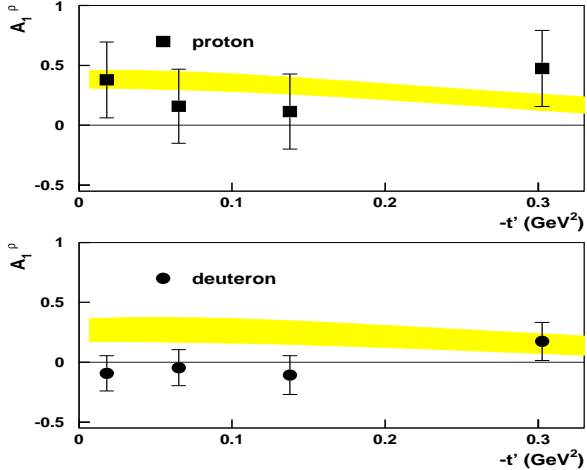}}
\caption{\baselineskip=0.38cm The $-t'$-dependence of the asymmetry $A^\rho_1$ in exclusive 
$\rho^0$ meson electroproduction on the proton (top) and deuteron (bottom). Error bars have the 
same meaning as in Fig. \ref{fig:3}. The shaded areas represent the range allowed for the 
theoretical predictions of Ref.~\cite{koch_new}.}
\label{fig:4}   
\end{figure}

\begin{figure}[t!]
\vspace*{-0.2cm}
\resizebox{0.5\textwidth}{0.40\textwidth}{\includegraphics{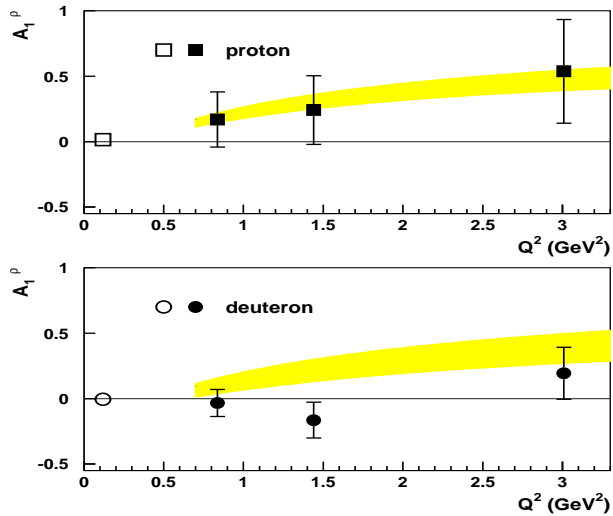}}
\caption{\baselineskip=0.38cm The $Q^2$-dependence of the asymmetry $A^\rho_{1}$ in 
exclusive electroproduction (closed symbols) and electroproduction by quasi-real photons 
(open symbols) on the proton (top) and deuteron (bottom). The error bars have the same meaning as 
in Fig. \ref{fig:3}. The uncertainties of the data from electroproduction by quasi-real photons are covered by the symbols. The shaded areas represent the range allowed for the 
theoretical predictions of Ref.~\cite{koch_new}.}
\label{fig:5}
\vspace*{-0.3cm}   
\end{figure}

The dependences of $A_1^\rho$ on the momentum transfer $-t'$ and $Q^2$ are
respectively shown in Fig. \ref{fig:4} and Fig. \ref{fig:5} for both the proton and deuteron. 
They are compared to theoretical predictions calculated recently in the framework of the Regge 
model~\cite{koch_new}. In this approach the parameters of the Reggeons exchanges
contributing to $\rho^0$ meson electroproduction on the nucleon were extracted from fits to 
the nucleon structure functions $g^N_1$ and $F^N_2$. These parameters were subsequently used to calculate
the $\rho^0$ electroproduction amplitudes with natural and unnatural parities. Sizeable double-spin
asymmetries are predicted for exclusive electroproduction on both proton and deuteron. 
While the predicted values are consistent with the measured ones on the 
proton, they are larger in the case of the deuteron. 
 
The double-spin asymmetries in $\rho^0$ electroproduction by quasi-real photons are also shown in 
Fig. \ref{fig:5} at the correspondingly low value of $Q^2$. They are consistent with zero for both 
the proton and deuteron.

\begin{figure}[t!]
\vspace*{-1.cm}
\resizebox{0.50\textwidth}{0.37\textwidth}{\includegraphics{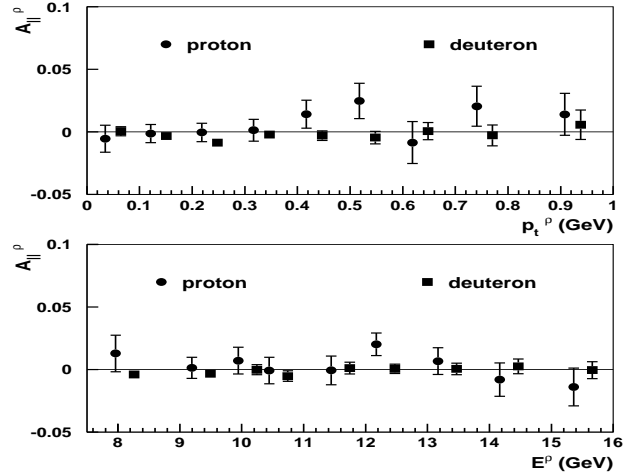}}
\caption{\baselineskip=0.38cm The dependence on the transverse momentum (top), calculated 
with respect to the beam direction, and energy of the $\rho^0$ (bottom) of the asymmetry $A^\rho_{||}$ in $\rho^0$ electroproduction by quasi-real photons on 
the proton (circles) and deuteron (squares). The proton 
data are slightly shifted to the left for clearer representation. Error bars have the 
same meaning as in Fig.\ref{fig:3}.}
\label{fig:6}
\vspace*{-0.3cm}   
\end{figure}

Lepton-nucleon asymmetries in $\rho^0$ electroproduction by quasi-real photons were 
measured over a range of values of $\rho^0$ meson energy and transverse momentum calculated with 
respect to the beam direction. As can be seen in Fig.~\ref{fig:6}, no trend was observed in any of these variables. 
Note that, as was already mentioned above, for electroproduction by quasi-real photons it is not 
possible to impose the requirement of exclusivity.

Double-spin asymmetries in exclusive $\rho^0$ meson electroproduction on the proton and 
deuteron were measured at SMC~\cite{tripet} at 
$\langle W \rangle $=15\,GeV. In this region Pomeron exchange is thought to dominate~\cite{gvmd} 
vector-meson production. In Ref.~\cite{tripet} the asymmetries were found to be consistent with 
zero for both the proton and deuteron in the region $0.01<Q^2<5$\,GeV$^2$, and slightly negative at 
$Q^2\sim$10\,GeV$^2$. This supports the expectation that Pomeron exchange is dominant at higher energies. 
In contrast, the tendency for a non-zero double-spin asymmetry found on the proton at 
HERMES energy suggests a significant contribution of the exchange of Reggeons or di-quark objects 
to the transverse part of exclusive $\rho^0$ electroproduction.  The asymmetry on the deuteron was 
measured to be consistent with zero. As unnatural parity exchange need not necessarily produce a 
non-zero asymmetry on all targets, this does not contradict the conclusion based on that result from 
the proton target.

In the case of $\phi$ meson electroproduction, the photoabsorption asymmetries are found 
to be consistent with zero in both event topologies considered here. 
A theoretical prediction exists for the case of electroproduction by quasi-real photons.
It implies sensitivity of the asymmetry to the strangeness 
content of the nucleon through interference of $s\bar{s}$
knockout with the diffractive VMD amplitude. In kinematic 
conditions ($\langle W\rangle=4.2$\,GeV, 
$\langle Q^2\rangle\simeq 0)$ similar to those of HERMES, 
and assuming the strange\-ness probability for the proton 
is $P_{s\bar{s}}=0.01$, the predicted asymmetry~\cite{titov} 
at $-t'$=0 ranges between -0.05 and +0.03, while for $-t'$=0.5 GeV$^2$ 
the range is -0.06 to +0.15, depending on the unknown relative phase 
of the amplitudes. The experimental result is compatible with zero 
strangeness content, but favours a negative phase $\delta_{s\bar{s}}$ 
if the strangeness is non-zero as assumed.

\section{ Summary}

Double-spin asymmetries in the cross section  of $\rho^0$ and $\phi$ electroproduction 
were measured by scattering longitudinally polarised leptons off longitudinally polarised hydrogen 
and deuterium targets at HERMES. The analysis was performed for two different event topologies: 
exclusive electroproduction, and electroproduction by quasi-real photons without the requirement 
of exclusivity.

The statistically weak evidence of a non-zero double-spin asymmetry in exclusive $\rho^0$ meson
electroproduction on the proton, as reported in Ref.~\cite{hermes_asym}, is also seen with the  
improved data set and analysis scheme. This suggests a contribution of unnatural-parity exchange 
to exclusive $\rho^0$ electroproduction by transverse photons at HERMES energies. 
An essentially flat dependence of the proton asymmetry on $t'$ is again observed, consistent with 
a prediction based on the description of the nucleon structure
functions in the framework of the Regge model~\cite{koch_new}. 

The same double-spin asymmetry  measured on the deu\-teron is found to be consistent with zero,
which disagrees with the prediction of Ref.~\cite{koch_new}.
The observed difference between the asymmetries measured on the proton and deuteron is consistent, 
however, with the expectation of Refs.~\cite{fraas,hermes_asym} which relates the asymmetries 
in $\rho^0$ production to those in inclusive DIS. 
 
The tendency towards a non-zero asymmetry found in exclusive $\rho^0$ electroproduction on the proton 
at HERMES, where quark exchange is expected to contribute substantially~\cite{hermes_cross}, can 
be reconciled with the zero asymmetry measured at the higher energy of the SMC experiment, 
where Pomeron exchange dominates and therefore no asymmetry in $\rho^0$ production is expected.

The double-spin asymmetry in $\rho^0$ electroproduction by quasi-real photons was found 
to be consistent with zero.

The measured asymmetries for $\phi$ mesons are consistent with zero within experimental uncertainties, 
both in exclusive electroproduction and electroproduction by quasi-real 
photons. This is consistent with the expected dominance of Pomeron exchange 
in $\phi$ electroproduction, and, in the case of electroproduction by quasi-real photons, with a theoretical prediction~\cite{titov} which involves 
$s\bar{s}$ knockout from the nucleon.

\begin{acknowledgement}
\baselineskip=0.38cm
We gratefully acknowledge the DESY management for its support and
the DESY staff and the staffs of the collaborating institutions.
This work was supported by
the FWO-Flanders, Belgium;
the Natural Sciences and Engineering Research Council of Canada;
the INTAS contribution from the European Commission;
the European Commission IHP program under contract HPRN-CT-2000-00130;
the German Bundes\-ministe\-rium f\"ur Bildung und Forschung (BMBF);
the Italian Istituto Na\-zio\-nale di Fisica Nucleare (INFN);
Monbusho International Scientific Research Program, JSPS, and Toray
Science Foundation of Japan;
the Dutch Foundation for Fundamenteel Onderzoek der Materie (FOM);
the U.K. Particle Physics and Astronomy Research Council and Engineering and
Physical Sciences Research Council; and
the U.S. Department of Energy and National Science Foundation.
\end{acknowledgement}

\end{document}